\documentclass[11pt,a4paper]{article}

\usepackage{amsfonts}
\usepackage{graphicx}

\title{\textbf{Kink dynamics in the MSTB Model}}

\author{A. Alonso Izquierdo$^{(a)}$
\\ {\normalsize {\it $^{(a)}$ Departamento de Matematica
Aplicada}, {\it Universidad de Salamanca, SPAIN}} }

\date{}

\begin{document}

\maketitle

\begin{abstract}
In this paper kink scattering processes are investigated in the Montonen-Sarker-Trullinger-Bishop model.  The MSTB model is in fact a one-parametric family of relativistic scalar field theories living in a one-time one-space Minkowski space-time which encompasses two coupled scalar fields. Between the static solutions of the model two kinds of topological kinks are distinguished in a precise range of the family parameter. In that regime there exists  one unstable  kink exhibiting only one non-null component of the scalar field. Another type of topological  kink solutions, stable in this case, includes two different kinks for which the two-components of the scalar field are  non-null.  Both one-component and two-component topological kinks are accompanied by their antikink partner. The  decay of  disintegration of the unstable kink to one of the stable pair plus radiation is  numerically computed. The pair of stable two-component kinks living respectively on  upper and lower half-ellipses in field space  belong to identical  topological sectors  in configuration space and  provides an ideal playground  to address several scattering events involving one kink and either its own antikinks or either the antikink of the other stable kink of the pair. By means of a numerical computation procedure we shall find and describe  interesting physical phenomena. Bion (kink-antikink oscillations) formation, kink reflection, kink-antikink annihilation, kink transmutation and resonances are examples of these type of events. The appearance of these special phenomena emerging in kink-antikink scattering configurations depends critically on the initial collision velocity and the chosen value of the coupling constant parametrizing the family of MSTB models.
\end{abstract}

\section{Introduction}

Over the last fifty years, topological defects behaving as solitary waves in non-linear scalar  field theories,  but  never occurring in linear system,  have been understood as the cornerstone in explaining the existence and the role of wall and/or brane structures in Condensed Matter \cite{Eschenfelder1981, Jona1993}, Cosmology \cite{Vilenkin1994}, Optics \cite{Agrawall1995}, Molecular systems  \cite{Davydov1985}, etc.. One-dimensional solitons or kinks, becoming domain walls in 3D space, are accompanied in different nonlinear gauge theories or sigma models by the existence of vortices and cosmic strings as line topological defects, monopoles and skyrmions, point defects, and instantons or textures, (-1)-brane defects, all of them sharing the essential feature of living in non-linear scenarios. We shall focus in this paper on solitons and kinks, whose paradigms are the solitary waves arising in the sine-Gordon and $\phi^4$ models. The impact of its study has been enormous both in the physical and mathematical literature in diverse contexts, despite that these models involve only one real scalar field. Search for static kinks in $N$-scalar field theories proved to be also an active research area, see, for instance, the references \cite{Rajaraman1982, Bazeia1995, Shifman1998, Alonso2002}. The discovery of kink solutions for which $N$-components of the scalar field arranged in an iso-vector were non-null opened a window to new possibilities ranging from its use to get a better knowledge of known phenomena to its application to understand another physical properties. In this paper we shall deal with the particularly interesting one-parametric family of relativistic (1+1)-dimensional $N=2$ scalar field theory known as the MSTB model. This system arises as a deformation of the $O(2)$ linear sigma model and has been the focus of study by many researchers for decades. It constitutes a natural generalization of the $\phi^4$ model, whose potential  further presents two absolute minima of the potential as a function of the two scalar field. Henceforth, the existence of two degenerate vacua and the associated spontaneous symmetry breakdown in the quantum version of the model is envisaged. The part of the potential energy density independent of the field derivatives in this system is the fourth-degree polynomial isotropic in quartic field powers but anisotropic in quadratic powers: $U(\phi_1,\phi_2)=\frac{1}{2}(\phi_1^2+\phi_2^2-1)^2 + \frac{1}{2}\sigma^2 \phi_2^2$. The anisotropy parameter $\sigma^2$ is the family parameter. A brief chronology of the main works concerning the family of MSTB models is organized in the following steps:

\begin{enumerate}
\item \textit{Birth of the model:} In 1976 Montonen discovered this family of models in his search for charged solitons in complex scalar field theories with a global $U(1)$ phase symmetry \cite{Montonen1976}. In his paper two different classes of static topological kinks were identified in the parameter range $\sigma\in (0,1)$: there were \textit{one-null component kinks}, for which the second scalar field component $\phi_2$ vanished whereas the $\phi_1$ kink profile is precisely the same as the kink profile in the standard $N=1$ $\phi^4$-model. There exist also \textit{two non-null component kinks}, such that the $\phi_1$ and $\phi_2$ kink profiles are both non-null and constrained to live in one of the two half-ellipses: $\phi_1^2+\frac{1}{1-\sigma^2}\phi_2^2=1$ with $\phi_2>0$ or $\phi_2<0$, in field space. One year earlier, Rajaraman and Weinberg obtained the first type of these solitary waves and described the qualitative behavior of the second class in a more general family of models \cite{Rajaraman1975}.

\item \textit{Stability analysis of the topological kinks:} Since it was known that the topology of the configuration space played a crucial role in the existence and  stability of  kinks  in one field scalar field theory the question arouse: which class of the MSTB kinks belonging both to the same topological sector is stable?. The stability analysis of the MSTB topological kinks was addressed and established by Sarker, Trullinger and Bishop from an energetic point of view by the end of 1976. They concluded that the two component non-null kinks are stable whereas the famous $\phi^4$ one-component kink embedded in the MSTB model still is a static solution but  is unstable \cite{Trullinger1976} in the $N=2$ ambient space. Further stability analysis based on the nature of the small kink fluctuations were performed in 1979, see \cite{Currie1979}.

\item \textit{Discovery of non-topological kinks:} In the same year, a non-topological kink, for which the two field components were non-null, was discovered by Rajaraman \cite{Rajaraman1979} for the parameter value $\sigma=\frac{1}{2}$, whose orbit in field space is a circle. The kink profile of the non-topological kinks tends to the same vacua at the two ends of the spatial line. Along the two subsequent years Subbaswamy and Trullinger numerically found that this kink was a single member of a one-parametric family of non-topological kinks. They proved the existence of this family in the parameter range $\sigma\in (0,1)$ and showed that these solutions are unstable \cite{Subbaswamy1980,Subbaswamy1981}. In addition it was checked the compliance of the so called \textit{energy sum rule}: the total energy of the non-topological kinks is the sum of the energies of the two classes of topological kinks.

\item \textit{Integrability of the analogue mechanical system:} In 1984, Magyari and Thomas \cite{Magyari1984} showed that the system of static field equations, equivalent to the Newton equations in the potential $V=-U$, is completely integrable by finding two constants of motion in involution  for the analogue mechanical system of two degrees of freedom. Indeed, the system is not only completely integrable but Hamilton-Jacobi separable by using elliptic coordinates. In 1985 Ito was able to obtain implicit expressions in these coordinates for every orbit in the whole static kink variety \cite{Ito1985}. He also proved that the non-topological kinks are unstable by applying the Morse index theorem to the kink orbit manifold \cite{Ito1985b}. This conclusion is based on the fact that all the non-topological kink orbits cross each other at one of the foci of the elliptic coordinate lines. In a series of three papers, \cite{Guilarte1987,Guilarte1988, Guilarte1992}, the  full Morse Theory of the MSTB configuration space was developed by  Mateos-Guilarte  through the understanding of  the kink variety in the MSTB model as the space of geodesics of the Maupertuis-Jacobi action of the analogue mechanical system.

\item \textit{Generalizations of the MSTB models:} In 1998 it was noted that the MSTB model is not a rara avis between relativistic two scalar field models. New two-component scalar field theory models, having Hamilton-Jacobi separable analogue mechanical system as well as rich varieties of kink orbits, were proposed and studied  in \cite{Alonso1998}. In 2000 extensions of the MSTB model to $N$-component scalar field theories, analogue mechanical systems with $N$ degrees of freedom,  were constructed and discussed in \cite{Alonso2000}. All these extensions are deformations of the $O(N)$ linear sigma model where the potential energy density remains being a fourth-degree polynomial isotropic in quartic but anisotropic in quadratic powers of the fields. In this  last paper the entire static kink manifold is analytically identified as in the MSTB model  by using a system of elliptic coordinates. The stability analysis of these kinks was completed in \cite{Alonso2002c}. In 2008, a systematic classification of the two-component generalized MSTB models and the description of its static kink manifolds were established  in the work \cite{Alonso2008}.

\item \textit{Quantum kinks:} Finally, it is worth mentioning that the promotion of the MSTB model to the quantum realm has been considered in \cite{Alonso2002b}. In this work the semiclassical mass of the stable static topological kinks is computed by controlling the ultraviolet divergences  in the generalized zeta function regularizatio scheme.
\end{enumerate}

All the results achieved in the previously alluded works were  obtained within the analysis of static structures, like instantaneous pictures in a movie. The central theme in this paper is the understanding and description of the kink dynamics in the MSTB model. For example, the scattering between two two-component topological kinks will be one of the main problems to be studied and will be thoroughly discussed. Pursuing this endeavour we shall encounter a great difficulty. Contrarily to the analogue mechanical system governing static solutions in the MSTB model search for MSTB solutions evolving in time, besides been spread along the spatial line is not an integrable problem in (1+1)D scalar field theory. The MSTB field theory is a non-integrable field theoretical system rather different to the integrable sine-Gordon field theory which admits an infinite number of conserved charges. The consequence is that we cannot apply analytical tools to study the dynamics of any object, extended or not, in the MSTB field theory. Therefore, we shall rely in our analysis on a mixture of numerical and symbolic computations.

Rather than meson scattering we are interested in the study of kink-kink scattering in the MSTB model giving rise to very intriguing and complex  dynamical process. Collisions of infinitely extended objects may bring us to contemplate highly non-trivial and exotic evolution patterns. Kink-kink and kink-antikink collisions have been deeply studied in one-component scalar field theoretical models. Indeed, this subject drew great attention towards the seminal paper by Campbell and collaborators \cite{Campbell1983}. In this work, Campbell, Schonfeld and Wingate investigated the dynamical interactions between kinks and antikinks in the archetypical $\phi^4$ model by varying initial collision velocities. For initial collision velocity greater than a critical velocity $v_c\approx 0.2598$ kink reflection takes place. If the initial collision velocity is great enough the kink and the antikink collide, bounce back and escape respectively towards $x=-\infty$ and $x=+\infty$ losing certain amount of energy through meson radiation emission. If $v_0<v_c$, however, the kink and the partner antikink are compelled to collide a second time. In fact the formation of a kink-antikink quasi-bound state, a bion, is prevalent in this range $v_0<v_c$. The kink and the antikink collide and bounce back over and over again, losing a decreasing amount of kinetic energy in every impact. Moreover, there exist certain initial velocity windows in this regime where the kink and the antikink escape after the second impact. Narrower velocity windows were also found in the $\phi^4$-model where the kink and the antikink escape after colliding $N\geq 3$ times. Campbell and his collaborators were able to explain this behaviour by using the collective coordinates approach initially introduced for the $\phi^4$ model in \cite{Sugiyama1979} and later corrected in \cite{Weigel2014, Takyi2016}. These authors concluded that the so called resonant energy transfer mechanism is responsible for this phenomenon. In this process there is an energy exchange between the kink translational mode and the internal vibrational mode in each collision. Another novel property unveiled in this work is that the distribution of the resonant windows exhibits a fractal structure \cite{Anninos1991}. An analytical explanation of this phenomenon, based on a collective coordinates model for the resonant energy transfer mechanism, can be found in references \cite{Goodman2008,Goodman2005,Goodman2007}. Similar patterns have also been found in many other one-component scalar field theories. For example, the kink-antikink scattering has been investigated in the modified sine-Gordon model \cite{Peyrard1983}, in the $\phi^6$ model \cite{Weigel2014, Gani2014}, in the $\phi^8$ model \cite{Belendryasova2017}, in non-polynomial models \cite{Bazeia2017a,Bazeia2017b}, etc. All these studies have led to the conclusion that the relationship between the resonant energy transfer mechanism and the kink vibrational modes is more complicated than previously thought. Indeed, the kink in the $\phi^6$ model lacks internal vibrational modes \cite{Weigel2014} but the resonant energy transfer mechanism operating in this model is triggered by an internal vibrational mode of the combined kink-antikink configuration \cite{Dorey2011}. On the other hand, the existence of many kink vibrational modes can provoke the suppression of bounce-windows in kink-antikink collisions \cite{Simas2016} or the inclusion of quasiresonances \cite{Campbell1986,Gani1999,Gani2018}. In a recent paper \cite{Dorey2018} Dorey and Romanczukiewicz have demonstrated that the presence of quasi-normal modes can also serve as catalyst for the formation of resonance windows. In addition to the previous works, kink dynamics in one-component scalar field theory models that involve impurities, defects or local inhomogeneities have been considered in the references \cite{Fei1992, Fei1992b, Malomed1985, Malomed1989, Malomed1992, Goodman2004, Javidan2006, Saadatmand2012}. Another interesting phenomenon in this framework, known as the negative radiation pressure, is discussed in \cite{Romanczukiewicz2008, Romanczukiewicz2017}. In this situation a kink hit by a plane wave is accelerated towards the source of radiation. Finally, it is also worth to mention the investigation on collision of vector solitons in the coupled nonlinear Schr\"odinger model \cite{Tan2001,Yang2000} and the kink scattering in some two-component scalar field theories \cite{Halavanau2012, Alonso2017}.

We plan thus to investigate here similar phenomena arising in kink collision dynamics in the MSTB models. Besides, another kink evolution is worth to study in the MSTB model: the decay process of the one-component unstable topological kink or its antikink. We want thus to elucidate the mechanism of disintegration of the unstable kinks and the description of the emerging objects in the final state for this event. The kink manifold also includes two stable topological kinks together with their corresponding antikinks. It is crucial to understand and classify the possible scattering processes between these extended particles. A specific kink can be forced to collide with its own antikink or the antikink of the second existing stable topological kink. It is interesting to know if resonance phenomena arise in these cases. All of these events are explored in detail in this paper. Some partial results about the scattering between topological kinks with opposite topological charge but not forming an antikink-kink pair has been described in \cite{Alonso2018}.

The organization of this paper is as follows: in Section \S.2 the MSTB model is introduced and its variety of kink solitary waves is described; in Section \S.3 the kink dynamics in the MSTB model is thoroughly discussed, in particular the disintegration of the unstable kinks and the two types of kink/antikink scattering processes are explained. Finally, some conclusions are drawn and future prospects are outlined in Section \S.4.

\section{The MSTB model and its static kink variety}

The dynamics of the one-parameter family of MSTB models is governed by the action
\begin{equation}
S=\int d^2x \left[ \frac{1}{2}\partial_\mu \phi_a \partial^\mu \phi_a - U(\phi_1,\phi_2) \right] \hspace{0.3cm} , \label{action}
\end{equation}
where the functional $U[\phi_1,\phi_2]$ as a function of the fields  is the fourth-degree polynomial
\begin{equation}
U(\phi_1,\phi_2)= \frac{1}{2} \, (\phi_1^2+\phi_2^2-1)^2 + \frac{\sigma^2}{2} \, \phi_2^2 \hspace{0.3cm}. \label{potential}
\end{equation}
Here $\phi^a : \mathbb{R}^{1,1} \rightarrow \mathbb{R}$, $a=1,2$, are two dimensionless real scalar fields and the Minkowski metric $g_{\mu\nu}$ is chosen as $g_{00}=-g_{11}=1$ and $g_{12}=g_{21}=0$. The notation $x^0\equiv t$ and $x^1\equiv x$ will be used from now on. The coupling constant $\sigma$ arising in the second summand of (\ref{potential}) is a real parameter, $\sigma \in \mathbb{R}$. The MSTB model is thus a deformation of the $O(2)$-linear sigma model, where explicit symmetry breaking of $O(2)$ to the discrete sub-group $\mathbb{Z}_2\times\mathbb{Z}_2$ generated respectively by  $(\phi_1\to -\phi_1, \phi_2\to \phi_2)$ and  $(\phi_1\to  \phi_1, \phi_2\to -\phi_2)$
 takes place due to the last summand in (\ref{potential}).

The system of coupled  PDE equations
\begin{eqnarray}
\frac{\partial^2 \phi_1}{\partial t^2} - \frac{\partial^2 \phi_1}{\partial x^2} &=& 2 \phi_1(1-\phi_1^2-\phi_2^2)  \hspace{1.1cm} , \label{klein1}\\
\frac{\partial^2 \phi_2}{\partial t^2} - \frac{\partial^2 \phi_2}{\partial x^2} &=& 2 \phi_2 (1-\phi_1^2-\phi_2^2-{\textstyle\frac{\sigma^2}{2}}) \hspace{0.3cm}, \label{klein2}
\end{eqnarray}
encompasses the two Euler-Lagrange equations of the action functional (\ref{action}). The configuration space is defined as the set of finite energy maps from the Minkowski space-time to the field space, i.e., ${\cal C}= \{ \Phi(x,t)\equiv (\phi_1(x,t),\phi_2(x,t)) \in {\rm Maps}(\mathbb{R}^{1,1}, \mathbb{R}^2) : E[\Phi(x,t)] < +\infty\}$. The energy density carried by a particular configuration $\Phi(x,t)=(\phi_1(x,t),\phi_2(x,t))$ is :
\[
{\cal E} \left[\Phi(x,t)\right] = \frac{1}{2} \Big( \frac{\partial \phi_1}{\partial t} \Big)^2 + \frac{1}{2} \Big( \frac{\partial \phi_2}{\partial t} \Big)^2 +\frac{1}{2}\Big(\frac{\partial \phi_1}{\partial x} \Big)^2 + \frac{1}{2} \Big( \frac{\partial \phi_2}{\partial x} \Big)^2 + U(\phi_1,\phi_2) \hspace{0.3cm},
\]
whereas its spatial integration along the whole  real line defines the total energy:
\begin{equation}
E[\phi_1,\phi_2]=\int_{-\infty}^\infty \!\! dx \,\, {\cal E} \left[\Phi(x,t)\right] \hspace{0.2cm} . \label{totalenergy}
\end{equation}
Evolution of the different elements in the configuration space taken as initial values of the system (\ref{klein1})-(\ref{klein2}) is determined by
solving the corresponding Cauchy problems for this PDE system. The energy finiteness condition forces configurations to satisfy the following asymptotic conditions:
\[
\lim_{x\rightarrow \pm \infty} \frac{\partial \Phi(x,t)}{\partial t}= \lim_{x\rightarrow \pm \infty} \frac{\partial \Phi(x,t)}{\partial x}=  0 \hspace{0.8cm} \mbox{and} \hspace{0.8cm}
\lim_{x\rightarrow \pm \infty} \Phi(x,t)  \in {\cal M} \hspace{0.2cm} ,
\]
where ${\cal M}=\{A_+ =(+ 1,0), A_- =(-1,0)\}$ is the set of zeros or absolute minima of the MSTB potential $U(\phi_1,\phi_2)$. Since ${\cal M}$ is a discrete set $\Phi(\pm \infty,t)$ are constant of the motion because any variation of the asymptotic values of the fields would cost infinite energy. The configuration space is therefore the union of four disconnected topological sectors: ${\cal C}={\cal C}^{++} \cup {\cal C}^{+-}\cup {\cal C}^{-+}\cup {\cal C}^{--}$ distinguished by the four admissible values of the fields at the two ends of the real spatial line. It is standard to define the \lq\lq topological\rq\rq charge
\[
q = \frac{1}{2} \,  \Big( \phi_1(+\infty,t) - \phi_1(-\infty,t ) \Big) \hspace{0.2cm} ,
\]
as the invariant distinguishing between the different sectors of the configuration space{\footnote{To distinguish between ${\cal C}^{++}$ and ${\cal C}^{--}$, both having $q=0$, one needs to fix, e.g., $\phi_1(-\infty,t)$ also.}}. Configurations carrying non-zero topological charges
 living in  ${\cal C}^{+-}$, $q=+1$,  or ${\cal C}^{-+}$, $q=-1$, stay at their sector and are unable
to evolve in time to configurations belonging to ${\cal C}^{++}$ and  ${\cal C}^{--}$, all the forbidden evolutions would require infinite energy.

The simplest solutions of the partial differential equation system (\ref{klein1})-(\ref{klein2}) are static and homogeneous, precisely the elements of the set ${\cal M}$, which are the absolute minima of $U$: $\Phi_\pm (x,t)= A_\pm =(\pm 1, 0)$. Therefore, these zero energy solutions are classically stable and provide bona fide ground states to quantize the MSTB model: The choice of one of the two degenerate absolute minima of $U$ as the vacuum of the quantum version of the model spontaneously breaks further the remaining symmetry $\mathbb{Z}_2\times\mathbb{Z}_2$  to the $\mathbb{Z}_2$ sub-group generated by
$(\phi_1\to  \phi_1, \phi_2\to -\phi_2)$. The potential $U(\phi_1,\phi_2)$ as a function of the fields is non-negative and admits critical points that are thus static homogeneous solutions of the field equations. The character of the critical points depends on the ranges of $\sigma$. If $\sigma^2\in (0,2)$ the potential term $U(\phi_1,\phi_2)$ has two degenerate absolute minima $A_\pm =(\pm 1,0)$, a local maximum located at the internal plane origin $(0,0)$ and two saddle points placed at $(0,\sqrt{1-\sigma^2/2})$, see Figure 1(left). If $\sigma^2 \in [2,\infty)$ the potential term $U(\phi_1,\phi_2)$ has two absolute minima $A_\pm =(\pm 1,0)$ again, but now the origin $(0,0)$ becomes a saddle point of $U(\phi_1,\phi_2)$ and the saddle points of the previous range become imaginary losing their physical sense, see Figure 1(right). Only the absolute minima will play a role in the quantum realm because attempts to use the other types of classical solutions as quantum ground states will be plagued with tachyons in at least one of the two phonon branches.
\begin{figure}[h]
\centerline{\includegraphics[height=3.5cm]{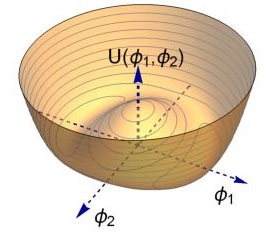} \hspace{2cm} \includegraphics[height=3.5cm]{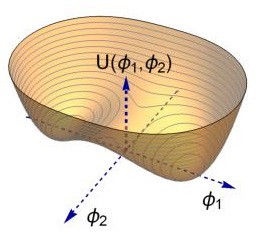}}
\caption{\small Graphical representation of the potential term (\ref{potential}) for $\sigma=0.5$ (left) and $\sigma=1.5$ (right). Notice that in the first case $(\phi_1,\phi_2)=(0,0)$ is a local maximum whereas in the second case is a saddle point.}
\end{figure}

The next step is the search for static but space-dependent solutions to the field equations (\ref{klein1}) and (\ref{klein2}) which become:
\begin{equation}
\frac{d^2 \phi_1}{d x^2} = -2 \phi_1(1-\phi_1^2-\phi_2^2)  \hspace{1.1cm} , \hspace{1cm}
\frac{d^2 \phi_2}{d x^2} = -2 \phi_2 (1-\phi_1^2-\phi_2^2-{\textstyle\frac{\sigma^2}{2}}) \hspace{0.3cm}. \label{newton}
\end{equation}
Re-interpreting $x$ as mechanical time and thinking of $(\phi_1,\phi_2)$ as the coordinates of a particle moving in a plane, the ODE system (\ref{newton}) is no more than the Newton equations for a particle moving in the force field created by the potential $V=-U$. Kinks, which are static solutions of the field equations of finite energy, or, localized  energy density, correspond in this way to finite mechanical action trajectories of this Newtonian system. It happens that mechanical systems with two degrees of freedom isotropic in quartic powers of the coordinates but anisotropic in the quadratic powers are Hamilton-Jacobi separable by using elliptic coordinates. In the Euler version the variable $u\in[\sigma, +\infty)$ measures half the sum of the distances of the particle position to two fixed points in the plane $F_\pm=(\pm\sigma,0)$ and $v\in[-\sigma,\sigma]$ is half the difference between these distances. The change of coordinates is tantamount to a map from the infinite strip to the upper half plane: $\rho_\pm :[\sigma,\infty)\times [-\sigma,\sigma] \rightarrow \mathbb{R}^2/\mathbb{Z}_2$. Allowing negative signs also in the $\phi_2$ coordinates the $1:2$ map reads:
\begin{equation}
\rho_\pm^* (\phi_1)\equiv \phi_1=\frac{1}{\sigma} u v \hspace{0.5cm}, \hspace{0.5cm} \rho_\pm^* (\phi_2)\equiv\phi_2 =\pm \frac{1}{\sigma} \sqrt{(u^2-\sigma^2)(\sigma^2-v^2)}  \hspace{0.3cm} .\label{elliptic}
\end{equation}
The energy functional (\ref{totalenergy}) expressed in these elliptic coordinates
\begin{eqnarray*}
E[\Phi(x)] &=& \int_{-\infty}^\infty dx \Big[ \frac{1}{2} \frac{u^2-v^2}{u^2-\sigma^2} \Big( \frac{du}{dx} \Big)^2 + \frac{1}{2} \frac{u^2-v^2}{\sigma^2-v^2} \Big( \frac{dv}{dx} \Big)^2 + \\ && \hspace{0.7cm} + \frac{1}{2(u^2-v^2)} [(u^2-1)^2(u^2-\sigma^2) + (1-v^2)^2 (\sigma^2-v^2)] \Big]
\end{eqnarray*}
may be rewritten \'a la Bogomolny in the form:
\begin{equation}
E[\Phi(x)] = \int_{-\infty}^\infty \!\!\! dx \Big\{ \frac{1}{2} \frac{u^2-v^2}{u^2-\sigma^2} \Big[ \frac{du}{dx} -(-1)^a \frac{u^2-\sigma^2}{u^2-v^2} (1-u^2) \Big]^2 + \frac{1}{2} \frac{u^2-v^2}{\sigma^2-v^2} \Big[ \frac{dv}{dx} -(-1)^b \frac{\sigma^2-v^2}{u^2-v^2} (1-v^2) \Big]^2 \Big\} + |T| \label{bogoener}
\end{equation}
where $a,b=0,1$ and
\begin{equation}
|T| =   \int_{-\infty}^\infty dx \Big| \frac{du}{dx} (1-u^2) \Big| + \int_{-\infty}^\infty dx  \Big| \frac{dv}{dx} (1-v^2) \Big| \hspace{0.3cm} . \label{bound}
\end{equation}
Given the structure of the functional (\ref{bogoener}), one sees that the Bogomolny-Prasad-Sommerfeld bound $|T|$ is saturated by static configurations complying with any of the following four systems of first-order ODE's:
\begin{equation}
\frac{du}{dx} = (-1)^a \frac{u^2-\sigma^2}{u^2-v^2} (1-u^2)  \hspace{1cm},\hspace{1cm} \frac{dv}{dx} = (-1)^b \frac{\sigma^2-v^2}{u^2-v^2} (1-v^2)  \hspace{0.3cm} .\label{bpsequations}
\end{equation}
After finding the finite \lq\lq action\rq\rq solutions of the ODE system (\ref{bpsequations}) one immediately obtains all the kink solitary waves of the MSTB model by going back to Cartesian coordinates by means of the change of variables (\ref{elliptic}). The coordinate lines back in the $(\phi_1,\phi_2)$-plane are confocal ellipses and hyperbolae whose foci are located at $F_\pm =(\pm \sigma, \sigma)$. We remark that the two copies $\rho_\pm$ are needed in (\ref{elliptic}) to cover the entire plane. Thus, smoothness conditions must be imposed on the solutions when crossing the axis $\phi_1$. The static kink variety of the MSTB model will be analytically identified on these grounds.

Before of doing that it is convenient to distinguish two Regimes determined from the $\sigma$ parameter where different kink patterns arise: (1) \textit{Regime A}: $\sigma\in[0,1)$ and (2) \textit{Regime B}: $\sigma\in[1,+\infty)$. We start now describing the topological kinks:

\vspace{0.1cm}

\noindent -- (1) \textit{One non-null component topological kinks} ${\cal K}_{\rm static}^{(q)}(x)$. For  any positive value of $\sigma$ kink solutions whose second field component vanishes exist. In this case the kink solutions are given by
\begin{equation}
{\cal K}_{\rm static}^{(q)}(x)=(q \tanh \overline{x},0) \hspace{0.3cm}, \label{kink1}
\end{equation}
where $\overline{x}=x-x_0$ with $x_0\in \mathbb{R}$ being the kink center. Here $q=\pm 1$ is the topological charge which distinguishes respectively between kinks and antikinks. Notice that the mirror symmetry $\pi_x:x\mapsto -x$ relates these solutions. Obviously, the minima $A_+$ and $A_-$ are connected by these kinks by means of the straight line $\phi_2=0$, see Figure 2. The energy density of these solutions is depicted also in Figure 2. These topological defects consist of only one energy density lump, thus, they may be interpreted as basic extended particles of the physical system.

\begin{figure}[h]
\centerline{\includegraphics[height=2.7cm]{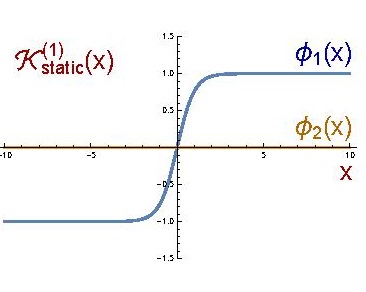} \hspace{0.2cm} \includegraphics[height=2.7cm]{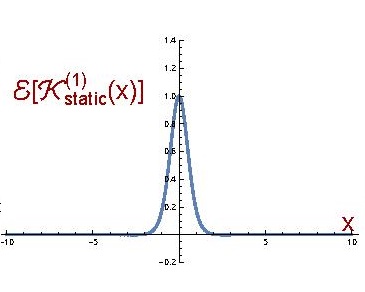} \hspace{0.2cm} \includegraphics[height=2.7cm]{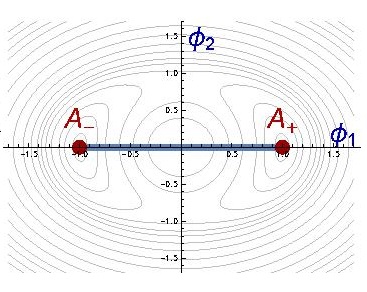}\hspace{0.2cm} \includegraphics[height=2.7cm]{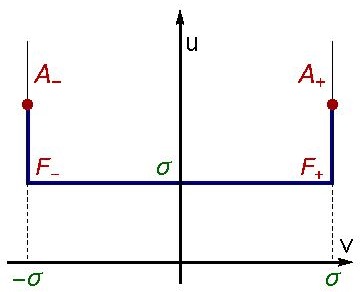}}
\caption{\small Graphical representations of the field components (a), energy density (b), and the orbits in the Cartesian (c) and elliptic (d) plane for the ${\cal K}_{\rm static}^{(q)}(x)$-kink.}
\end{figure}

\noindent The description of these kinks in the elliptic plane comprises two possibilities:

\begin{enumerate}
\item In the \textit{regime B} ($\sigma>1$) the vacuum points are characterized by $u_\pm=\sigma$ and $v_\pm=\pm 1$. The condition $u=\sigma$ solves the $u$-equation in (\ref{bpsequations}). The ensuing $v$-equation $\frac{dv}{dx}=(-1)^a (1-v^2)$ is easily integrated in the range $-1<v<1$ to find $v(x)=(-1)^a \tanh (x-x_0)$. These expression leads to the solution (\ref{kink1}). In this case, the kink energy is a proper topological bound
\[
E[{\cal K}_{\rm static}^{(q)}(x)]= |T| =  \int_{-1}^1 dv \, (1-v^2) = \frac{4}{3} \, \, \, \hspace{0.3cm} .
\]

\item In the \textit{Regime A} ($0<\sigma<1$) the vacuum points are $u_\pm=1$ and $v_\pm=\pm \sigma$. One non-null component kinks also exist in this case but are composed of three steps in the $(u,v)$ strip. In the first stage $u$ varies in the range $\sigma<u<1$ but $v=-\sigma$ remains fixed. One must integrate $\frac{d u}{dx}=(-1)^a(1-u^2)$ to find $u(x)=(-1)^a\tanh (x-x_0)$ for $x-x_0\in (-\infty,{\rm arctanh} \, \sigma)$. The second stage runs with $u=\sigma$ being constant and $v$ varying in the range $-\sigma<v<\sigma$ according to the equation: $\frac{dv}{dx}=(-1)^a(1-v^2)$. The trajectory is therefore: $v(x)=(-1)^a\tanh(x-x_0)$, starting and ending at the foci: $-{\rm arctanh}\,\sigma < x-x_0 < {\rm arctanh}\, \sigma$ if $a=0$. The inequality goes in the other sense
if $a=1$. The third stage is the reverse of the first stage although now $v=\sigma$ remains constant and $u$ varies in the interval $\sigma<u<1$. The trajectory is $u(x)=(-1)^a\tanh(x-x_0)$. Back in Cartesian coordinates the kink solitary wave if $\sigma<1$ follows the form (\ref{kink1}). In \textit{Regime A} the one non-null component kink energy is not a proper topological quantity because it depends on two points in the mid of the trajectory:
\[
E[{\cal K}_{\rm static}^{(q)}(x)]= 2 \int_\sigma^1 dv (1-v^2) + \int_{-\sigma}^\sigma du (1-u^2) = \frac{4}{3} \, \, \, .
\]
In \textit{Regime A} one non-null component kinks are not BPS states.
\end{enumerate}

Thus, even though being apparently identical, one non-null component topological kinks are very different in \textit{Regimes A} and \textit{B}. In the original field variables the difference emerges in the study of the stability of these solutions.
Linear stability of a static solution $\Phi(x)$ is dictated by the evolution of small fluctuations around the solution $\Phi(x)$. In this context, it is imposed that the perturbed solution $\Psi(x,t)=\Phi (x) + \epsilon e^{i\omega t} F^\omega(x)$ be a solution of field equations (\ref{klein1}) and (\ref{klein2}) up to first order in the infinitesimal parameter $\epsilon$. As a result, the two-component perturbations $F^\omega(x) =(f_1^\omega(x),f_2^\omega(x))^t$ must be eigenfunctions of the second order small fluctuation matrix operator
\[
{\cal H}_{ij}[\Phi(x)] =  - \delta_{ij} \frac{d^2}{dx^2} + V_{ij}(x)= - \delta_{ij} \frac{d^2}{dx^2} + \frac{\partial^2 U}{\partial \phi_i\partial \phi_j} [\Phi(x)] \hspace{0.4cm},\hspace{0.4cm} i,j=1,2 \hspace{0.3cm} ,
\]
belonging to a rigged Hilbert space $H$. In other words, the spectral equation
\begin{equation}
{\cal H}[\Phi(x)] F^\omega (x)= \omega^2 F^\omega(x) \label{spectraleq}
\end{equation}
holds if $\Psi(x,t)$ is still solution and $\Phi(x)$ is stable if $\omega^2>0$, although neutral equilibrium small fluctuations may exist for which $\omega^2=0$. Analytical identification of the spectrum of the matrix operator ${\cal H}$, ${\rm Spec}({\cal H})=\{\omega^2 \in \mathbb{R}: (\exists F^\omega(x) : {\cal H}F^\omega(x)=\omega^2 F^\omega(x),F^\omega(x)\in H)\}$ is, in general, unapproachable. However, the ${\cal K}_{\rm static}^{(q)}(x)$-kink fluctuation operator
\[
{\cal H}[{\cal K}^{(q)}(\overline{x})]= \left( \begin{array}{cc} -\frac{d^2}{dx^2} + 4- 6 \, {\rm sech}^2 \overline{x} & 0 \\ 0 & - \frac{d^2}{dx^2} + \sigma^2 - 2 \, {\rm sech}^2 \overline{x} \end{array} \right)
\]
is diagonal and the spectral problem in this case corresponds to two exactly solvable spectral problems  (independents from each other) for Schr\"odinger operators with transparent P\"oschl-Teller potentials. The longitudinal eigenmodes $F_1^\omega=(f_1^\omega(x),0)^t$ comprise a zero mode $F_1^0(x)=({\rm sech}^2 x,0)^t$, an excited discrete eigenmode $F_1^{\sqrt{3}}(x)=({\rm sech} x \tanh x,0)$ with eigenvalue $(\omega_1)^2=3$ and a continuous spectrum which emerges on the threshold value 4, $\omega_{k_1}^2=4+k_1^2$ with $k_1\in \mathbb{R}$ and eigenfunctions which are plane waves times the second Jacobi polynomial in $\tanh x$. The discrete spectrum of the transverse fluctuations $F_2^\omega=(0, f_2^\omega(x))^t$ contains only the eigenvalue $(\omega_2)^2=\sigma^2-1$ with eigenfunction $F_2^{\omega_2}=(0,{\rm sech} \, x )^t$ while the continuous spectrum  $\omega_{k_2}^2=\sigma^2+k_2^2$ with $k_2\in \mathbb{R}$ is in this case build from similar eigenfunctions replacing the second by the first Jacobi polynomial. In Figure 3 the complete spectrum of the operator ${\cal H}[{\cal K}^{(q)}(\overline{x})]$ is depicted as a function of the coupling constant $\sigma$ in the interval $[0,3]$.
The spectra of both the longitudinal and the transverse kink fluctuation operators have been overlapped in Figure 3. Note that discrete longitudinal/tranverse eigenvalues can be immersed in the continuous spectrum of the tranverse/longitudinal fluctuations. The most relevant result from this analysis is that the eigenvalue $\omega_2^2=\sigma^2-1$ (emerging in the transverse fluctuation operator) is negative if $0<\sigma^2<1$. Therefore, the one-null component static kink ${\cal K}^{(q)}(x)$ is unstable in \textit{Regime} A and stable in Regime B. Indeed, this is the only topological defect solution which exists in \textit{Regime} B.

\begin{figure}[h]
\centerline{\includegraphics[height=3cm]{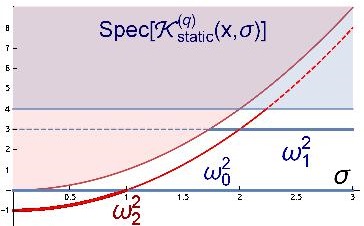}}
\caption{\small Spectrum of the small kink fluctuation operator ${\cal H}[{\cal K}^{(q)}(x)]$ as a function of the coupling constant $\sigma$.}
\end{figure}

\noindent -- (2) \textit{Two types of two non-null component topological kinks $K_{\rm static}^{(q,\lambda)}(x)$}: In \textit{Regime A} ($0<\sigma<1$), one non-null component topological kinks are unstable. There are, however, two classes of topological kinks which are stable.
In the $(u,v)$-strip one searches for trajectories where $u=1$ is fixed but $v$ varies in the interval $-\sigma<v<\sigma$ joining the two vacua in one stage. The solution of the ODE $\frac{dv}{dx}=(-1)^b (\sigma^2-v^2)$ leads to the expression $v=(-1)^b\sigma\tanh\sigma(x-x_0)$. Back in Cartesian coordinates in field space we find the two pairs of kink/antikinks
\begin{equation}
K_{\rm static}^{(q,\lambda)}(x)= \left(q \, \tanh (\sigma \overline{x}) , \lambda \sqrt{1-\sigma^2} \, {\rm sech} (\sigma \overline{x}) \right) \hspace{0.3cm}. \label{kink2}
\end{equation}
Here $q=\pm 1$ is the topological charge and $\lambda=\pm 1$ distinguishes if the second field $\phi_2$ is positive o negative, see Figure 4. It is immediate to check that the two components of these kinks live on an ellipse
\begin{equation}
\phi_1^2+ \frac{\phi_2^2}{1-\sigma^2} = 1 \hspace{0.3cm}
\label{ellipse}
\end{equation}
in field space, or better, the upper and lower half-ellipses depending on the sign of the second field component. Charge conjugation turns a kink into its antikink, i.e., $\overline{K}{}^{(q,\lambda)}(x)=K^{(q,\lambda)}(-x) = K^{(-q,\lambda)}(x)$.  Writing the energy of the $K_{\rm static}^{(q,\lambda)} (x)$-kinks as a topological BPS bound
\[
E[K_{\rm static}^{(q,\lambda)}(x)]= |T| =  \int_{-\sigma}^\sigma du \, (1-u^2) = 2\sigma \Big(1- \frac{\sigma^2}{3} \Big) \hspace{0.3cm} .
\]
the two non-null component topological kinks emerge as BPS states, a fact that ensures their absolute stability. These kinks may be interpreted as a basic single extended particle because the energy density is confined within a small region, see Figure 4.

\begin{figure}[h]
\centerline{\includegraphics[height=2.7cm]{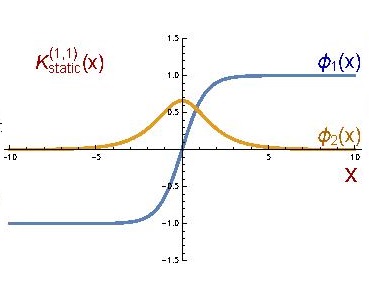} \hspace{0.2cm} \includegraphics[height=2.7cm]{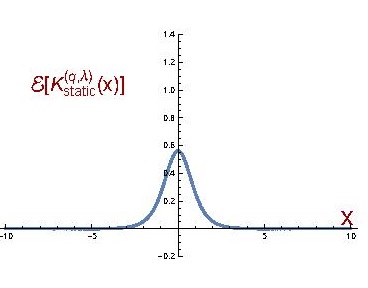} \hspace{0.2cm} \includegraphics[height=2.7cm]{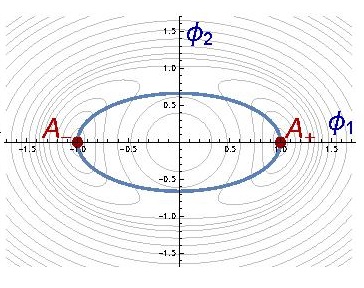} \hspace{0.2cm} \hspace{0.2cm} \includegraphics[height=2.7cm]{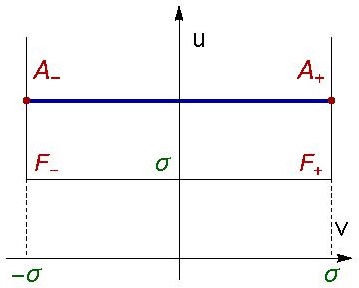}}
\caption{\small Graphical representations of the field components (a), energy density (b), and the orbits in the Cartesian (c) and elliptic (d) plane for the $K_{\rm static}^{(q,\lambda)} (x)$-kink.}
\end{figure}

\vspace{0.1cm}

The second-order small fluctuations around $K^{(q,\lambda)}(x)$-kinks are governed by the following $2\times 2$-matrix Schr$\ddot{\rm o}$dinger operator:
\begin{equation}
{\cal H}[K_{\rm static}^{(q,\lambda)}(x)]= \left( \begin{array}{cc} -\frac{d^2}{dx^2} + 4- 2(2+\sigma^2) \, {\rm sech}^2 \sigma \overline{x} & 4 \sqrt{1-\sigma^2} \, {\rm sech}\, \sigma \overline{x} \tanh \sigma \overline{x} \\  4 \sqrt{1-\sigma^2} \, {\rm sech}\, \sigma \overline{x} \tanh \sigma \overline{x} & - \frac{d^2}{dx^2} + \sigma^2 + 2 (2-3\sigma^2)\, {\rm sech}^2 \sigma \overline{x} \end{array} \right) \hspace{0.3cm} .
\label{hesstk2}
\end{equation}
Usually, one expands the small fluctuations in terms of the eigenfunctions of this matrix differential operator: ${\cal H}[K_{\rm static}^{(q,\lambda)}(x)] F^\nu(x)=\nu^2 F^\nu (x)$. No analytical information is available about this spectral problem except some qualitative features which guarantee that
these two types of two non-null topological kinks are stable in \textit{Regime A}. We mention the three main points: (1) The translational zero mode
\[
F^0(x)=\left({\rm sech}^2(\sigma x),\pm \overline{\sigma} \, {\rm sech}(\sigma x) \tanh (\sigma x) \right)^t
\]
belong to the kernel of ${\cal H}$ in the range $\sigma\in (0,1)$. (2) A pair of doubly degenerate continuous spectra emerging respectively on the threshold values $4$ and $\sigma^2$ exist. (3) In addition, numerical investigations reveal the presence of a discrete eigenvalue $\nu_1^2$ for large enough values of $\sigma$ in this \textit{Regime}. In Figure 5 the spectrum of the operator ${\cal H}[K_{\rm static}^{(q,\lambda)}(x)]$ is plotted for the range $\sigma\in (0,1]$. Observe that the discrete eigenvalue $\nu_1^2$ is non-negative. Thus, there are no negative eigenvalues in the spectrum of ${\cal H}[K_{\rm static}^{(q,\lambda)}(x)]$ and these kinks are stable, as previously pointed out.

\begin{figure}[h]
\centerline{\includegraphics[height=3cm]{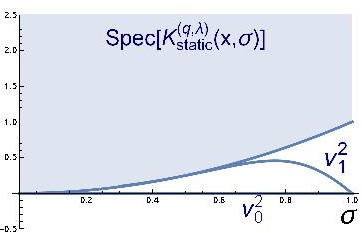}}
\caption{\small Spectrum of the small kink fluctuation operator ${\cal H}[K_{\rm static}^{(q,\lambda)} (x)]$ as a function of the coupling constant $\sigma$.}
\end{figure}

\noindent -- (3) \textit{Two one-parametric families of two non-null component non-topological kinks}: In \textit{Regime A}, there also exits a pair of one-parametric families of non-topological kinks $N_{\rm static}^\pm (x;\gamma)=(\phi_1^\pm (x,\gamma),\phi_2(x,\gamma))$, whose two field components are both non-null:
\begin{equation}
\phi_1^\pm (x;\gamma)= \pm \, \frac{\sigma_- \cosh (\sigma_+ x_+) - \sigma_+ \cosh (\sigma_- x_-) }{\sigma_- \cosh (\sigma_+x_+) + \sigma_+\cosh(\sigma_- x_-)}
\hspace{0.2cm}, \hspace{0.3cm}
\phi_2(x;\gamma)= \frac{2 \sigma_+ \sigma_ - \sinh \overline{x} }{\sigma_- \cosh (\sigma_+x_+) + \sigma_+\cosh (\sigma_- x_-)}
 \hspace{0.2cm}.
\label{mstbntk}
\end{equation}
The notation $\sigma_\pm = 1\pm \sigma$ and $x_\pm =\overline{x}-\gamma \sigma (\sigma \mp 1)$ have been used in (\ref{mstbntk}) to emphasize the regularity of these expressions. To derive the kink profiles (\ref{mstbntk})  the separability of the ODE system (\ref{bpsequations}) has been used. A particular member belonging to these families is singled out by  the value of the real parameter $\gamma\in \mathbb{R}$ and the $A_\pm$ asymptotic value reached by these kink trajectories at both ends of the spatial line. In Figure 6, the field components, the energy density and the orbit for the particular $N_{\rm static}^+ (x;\gamma)$-kink have been displayed for  $\gamma=6$. In general, all the $N_{\rm static}^\pm (x;\gamma)$-kinks connect one of  the two vacua $A_\pm$ with itself by means of closed orbits, all of them crossing one of the two foci $F_\mp$.

\vspace{0.2cm}

\begin{figure}[h]
\centerline{\includegraphics[width=3.5cm]{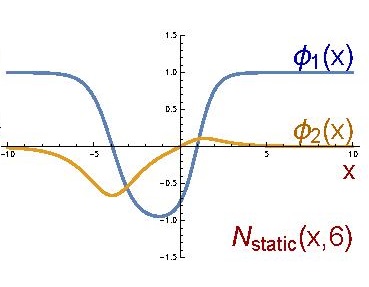} \hspace{0.2cm} \includegraphics[width=3.5cm]{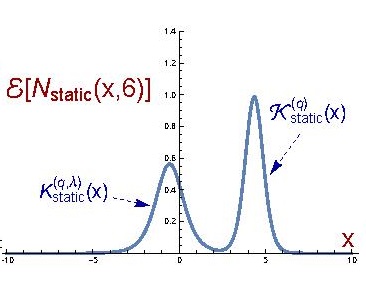} \hspace{0.2cm} \includegraphics[height=2.5cm]{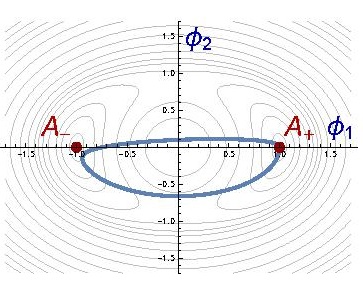} \hspace{0.2cm} \hspace{0.2cm} \includegraphics[height=2.5cm]{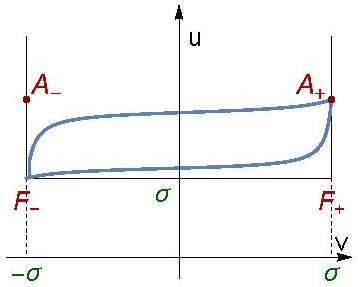}}
\caption{\small Graphical representations of the field components (a), energy density (b), and the orbits in the Cartesian (c) and elliptic (d) plane for the $N_{\rm static}^+(x;6)$-kink.}
\end{figure}

\vspace{0.2cm}

In the elliptic strip the projection of these kink trajectories runs twice over the $u\in [-\sigma,\sigma]$ and $v\in [\sigma,1]$ intervals, see Figure 6. Therefore, the following kink energy sum rule holds between the different types of MSTB kinks:
\[
E[N_{\rm static}^\pm (x;\gamma)]= 2 \int_\sigma^1 dv (1-v^2) + 2 \int_{-\sigma}^\sigma du (1-u^2) = E[K_{\rm static}^{(q,\lambda)}(x)]+E[{\cal K}_{\rm static}^{(q)}(x)]
\]
that is, the total energy of a non-topological kink is the sum of the total energies of the two classes of topological kinks. From the graphical representation of the $N_{\rm static}^\pm (x;\gamma)$-kink energy density in Figure 6, it can be understood that this relation is not accidental. Two energy lumps can be visualized in this graphics, which correspond to a $K_{\rm static}^{(q,\lambda)}(x)$ lump together a ${\cal K}_{\rm static}^{(q)} (x)$ lump. In other words, the $N_{\rm static}^\pm (x;\gamma)$-kinks describe a non-linear combination of the two basic extended particles of the system. The parameter $\gamma$ sets the separation between these particles, a kind of relative coordinate. For $\gamma=0$ the lumps are exactly overlapped whereas for large values of $\gamma$ the lumps of energy density are increasingly separated.

\section{Kink dynamics}

In this section the kink dynamics in the MSTB model will be numerically addressed. This study will be restricted to the Regime A where several types of kinks coexist. Two types of basic extended particles were identified in this regime, which are described by the topological $K_{\rm static}^{(q,\lambda)}(x)$ and ${\cal K}_{\rm static}^{(q)} (x)$ kinks. Note, however, that the second of these solutions is unstable, so the question about the fate of this unstable topological defect naturally arises. This particle carries a non-null topological charge and cannot decay to the vacuum sector. Therefore, the only possibility is that the unstable ${\cal K}_{\rm static}^{(q)} (x)$ kink decays into the stable $K_{\rm static}^{(q,\lambda)}(x)$ kink. This matter will be discussed in Section 3.1.

The scattering between two stable $K_{\rm static}^{(q,\lambda)}(x)$ kinks where $q,\lambda=\pm 1$ will also be investigated in this Section. Due to topological constraints the scattering processes must involve kinks with opposite topological charges. This allows the construction of a continuous initial multi-kink configuration, whose evolution is studied later. If symmetry considerations are also included in this framework, all the possible scattering events fall into one of the following two classes:
\begin{itemize}
\item[(a)] $K_{\rm static}^{(q,\lambda)}(x)- K_{\rm static}^{(-q,\lambda)}(x)$ scattering processes. The collisions between a kink-antikink pair are encompassed in this category. This kind of phenomena will be discussed in Section 3.2.

\item[(b)] $K_{\rm static}^{(q,\lambda)}(x)- K_{\rm static}^{(-q,-\lambda)}(x)$ scattering processes. These events comprise the collisions between a kink and the antikink of the other existing kink in this model. This situation will be described in Section 3.3.
\end{itemize}
As previously mentioned, numerical analysis is applied on the evolution equations (\ref{klein1}) and (\ref{klein2}) to determine the behavior of the scattering solutions. The numerical scheme that has been employed in this paper follows the algorithm introduced in \cite{Kassam2005} by Kassam and Trefethen, which is spectral in space and fourth order in time. As a complement to the previous scheme, an energy conservative second-order finite difference algorithm \cite{Strauss1978, Alonso2017} implemented with Mur boundary conditions \cite{Mur1981} has also been used. This algorithm lets to control the effect of radiation in the simulation because it absorbs the linear plane waves at the boundaries. The two previous numerical schemes provide identical results.

\subsection{Disintegration of the ${\cal K}^{(q)}(x)$-kink}

The linear stability study of the ${\cal K}^{(q)}(x)$-kink, given in Section 2, concludes that the application of an infinitesimal fluctuation following the form of the eigenmode $F_2^\omega(x)=(0,{\rm sech}\,x)^t$ of (\ref{hesstk2}) causes the instability of this solution. The evolution of the ${\cal K}^{(q)}(x)$-kink under these circumstances is investigated in this Section. Topological arguments maintains that this kink decays to the $K^{(q,\lambda)}(x)$ kink, which belongs to the same topological sector but is less energetic than the previous one.

In Figure 7 the evolution of the ${\cal K}^{(q)}(x)$-kink when slightly perturbed by a $F_2^\omega(x)$-fluctuation is displayed for the case $\sigma=0.7$. The two first graphics in Figure 7 illustrate the behavior of the field components. Globally, the initial configuration ${\cal K}^{(q)}(x)$ evolves to the $K^{(q,\lambda)}(x)$-kink although some internal vibration modes of this last solution are excited. Note, for example, the periodic oscillations of the maximum values of $\phi_2$ which are reached at $x=0$. A strong radiation emission is also apparent, mainly through the second field channel.

\vspace{0.4cm}

\begin{figure}[h]
\centerline{\includegraphics[height=3cm]{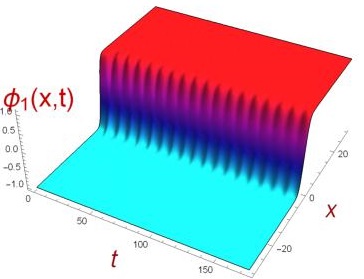} \hspace{0.2cm} \includegraphics[height=3cm]{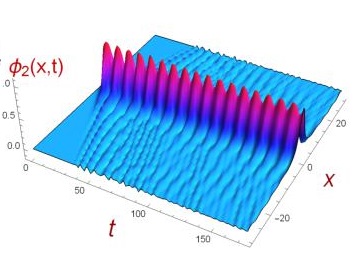} \hspace{0.8cm} \includegraphics[height=3cm]{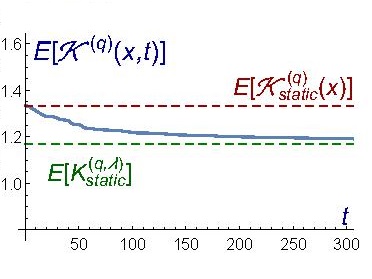} }
\caption{\small ${\cal K}^{(q)}(x)$-kink disintegration: Evolution of (a) the first field component, (b) the second field component and (c) the total energy of the evolving kink configuration (solid curve). The total energies of the static ${\cal K}^{(q)}(x)$ and $K^{(q,\lambda)}(x)$ kinks are depicted as dashed lines.}
\end{figure}

\vspace{0.2cm}

In Figure 7 the total energy of the evolving topological defect in the simulation interval is plotted by using a solid line. A large amount of the ${\cal K}^{(q)}(x,t)$-kink energy is lost due to radiation emission. This fact implies that $E[{\cal K}^{(q)}(x,t)]$ is a decreasing function as we can see in Figure 7(c). For the sake of comparison, the total energies of the static topological kinks ${\cal K}^{(q)}(x)$ and $K^{(q,\lambda)}(x)$ have been drawn by means of dashed lines in Figure 7. Observe that the ${\cal K}^{(q)}(x,t)$-kink total energy asymptotically approaches to the $K^{(q,\lambda)}(x)$-kink total energy, although a small amount of energy seems to be saved in internal vibrational eigenmodes. This process can be represented as
\[
{\cal K}^{(q)}(x) \rightarrow K^{*(q,\lambda)}(x) + \mbox{radiation}
\]
The asterisk superscript used in the previous relation emphasizes the fact that the resulting kink has excited internal vibration modes.

\begin{figure}[h]
\centerline{\includegraphics[height=4cm]{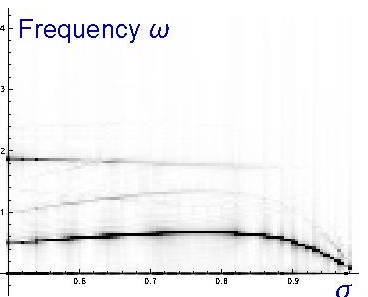} \hspace{1cm} \includegraphics[height=4cm]{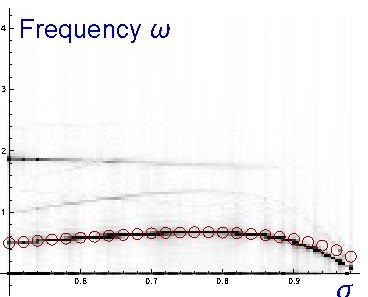} }
\caption{\small Spectral analysis of the second field component valued at the spatial origin $\phi_2(0,t)$ for the evolution of the static ${\cal K}^{(q)}(x)$-kink when perturbed by the fluctuation $F_2^\omega(x)$. In the second graphics the frequency $\nu_1$ of the operator ${\cal H}[K^{(q,\lambda)}(x)]$ have been overlapped as red small circles.}
\end{figure}

A explicit study of this fact can be visualized in Figure 8, where a spectral analysis of the behavior of the magnitude $\phi_2(0,t)$ is shown for values of the coupling constant $\sigma\in [0.5,1]$. The Fourier transform shows that several frequencies are excited with strengths that have been represented by using a gray scale. It can be checked that one of these frequencies is dominant over the other ones. In Figure 8(b) the square root of the eigenvalue $\nu_1^2$ of the second order small fluctuation operator (\ref{hesstk2}) has been plotted for several values of $\sigma$ overlapped with the previous spectral graphics. These frequencies of the operator ${\cal H}[K^{(q,\lambda)}(x)]$ are represented by small red circles. It can be checked the concordance between these values and the excited frequencies extracted from the spectral analysis. This fact allows to conclude that the ${\cal K}^{(q)}(x)$-kink decays to the $K^{(q,\lambda)}(x)$-kink and that this process excites the discrete eigen-fluctuation of this stable kink described in Section 2. This justifies the internal shape oscillations which suffer the $K^{(q,\lambda)}(x)$-kink after the disintegration, see Figure 7.

\subsection{$K_{\rm static}^{(q,\lambda)}(x)- K_{\rm static}^{(-q,\lambda)}(x)$ scattering processes}

The scattering between a stable kink $K^{(q,\lambda)}$ and its antikink $K^{(-q,\lambda)}$ is the topic numerically investigated in this Section. We are interested in classifying all the possible scattering events arising in this scenario, which depend on the initial velocity $v_0$ and the value of the coupling constant $\sigma$. The identities of the emerging topological defects and its separation velocity $v_f$ are the significant variables of this problem. The initial configuration for these numerical studies consists of two well separated boosted static kinks
\begin{equation}
K^{(q,\lambda)}(x-x_0,t;v_0) \cup K^{(-q,\lambda)}(x+x_0,t;-v_0) \label{conca2} \hspace{0.3cm} ,
\end{equation}
which are pushed together with speed $v_0$. Here $K^{(q,\lambda)}(x,t;v_0)=K_{\rm static}^{(q,\lambda)} [(x-v_0 t)/\sqrt{1-v_0^2} ]$. The trajectory of the multi-kink configuration (\ref{conca2}) describes a semi-elliptic curve, which is traversed twice. For $q=\pm 1$, the curve defined by (\ref{conca2}) goes from the vacuum $A_\mp$ to the opposite vacuum $A_\pm$ and later the same path is travelled in the reverse direction arriving to the point $A_\mp$ again. If $\lambda=1$ the multi-kink orbit (\ref{conca2}) lives in the semiplane $\phi_2\geq 0$ whereas if $\lambda=-1$ the second component of the concatenation (\ref{conca2}) is negative.

\begin{figure}[h]
\centerline{\begin{tabular}{c} \includegraphics[width=6cm]{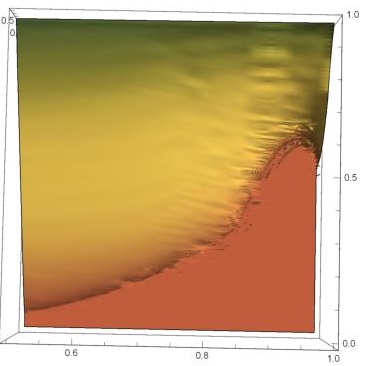} \end{tabular}\hspace{1cm}\begin{tabular}{c} \includegraphics[width=8cm]{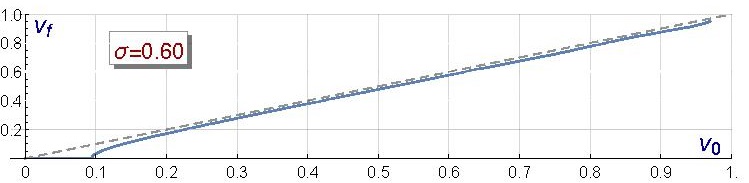} \\
\includegraphics[width=8cm]{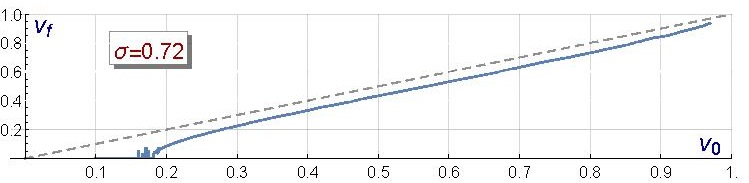} \\
\includegraphics[width=8cm]{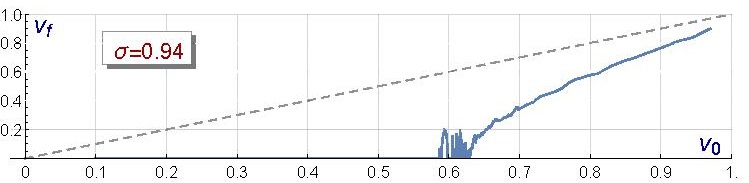} \end{tabular}}
\caption{\small Final kink velocity as a function of the initial velocity $v_0\in (0,1)$ and the model parameter $\sigma\in (0.5,1)$ for the $K^{(q,\lambda)}$-$K^{(-q,\lambda)}$ collisions (left, top view). Plane sections of the 3D graphics for the values $\sigma=0.6,0.71$ and $0.94$ (right). Zero final velocity indicates the formation of a kink-antikink bound state.}
\end{figure}

The final velocity $v_f$ of the scattered kinks is plotted in Figure 9 (left) as a function of the collision velocity $v_0$ and the coupling constant $\sigma$ where $\sigma\in [0.5,1)$. Plane sections of this 3D graphics for three fixed values of the parameter $\sigma$ are displayed in Figure 9 (right). The behaviour of these three plots are considered representative of the scattering events arising in this framework, which are described in the following points:

\vspace{0.2cm}

\noindent -- (1) For large enough values of the initial velocity $v_0$ the kink $K^{(q,\lambda)}(x)$ and its antikink $K^{(-q,\lambda)}(x)$ collide and emerge mutated into its $\phi_2$-mirror symmetric partners $K^{(q,-\lambda)}(x)$ and $K^{(-q,-\lambda)}(x)$, which travel away with a certain velocity $v_f<v_0$. A certain amount of kinetic energy is used to excite kink internal vibration eigenmodes and to emit radiation. This phenomenon, which is symbolized as
\begin{equation}
K^{(q,\lambda)}(v_0) + K^{(-q,\lambda)}(-v_0) \rightarrow K^{*(q,-\lambda)}(-v_f) + K^{*(-q,-\lambda)}(v_f) + {\rm radiation} \label{pheno1}
\end{equation}
is illustrated in Figure 10 for the particular values $\sigma=0.72$ and $v_0=0.5$. As before, the asterisk superscript stands for internal vibration mode excitation. The set of initial velocities $v_0$ and parameters $\sigma$ where this behaviour arises will be referred to as the \textit{one-bounce transmutation regime}. For a fixed value $\sigma$, the minimum initial velocity $v_c$ of this set will be named  \textit{critical velocity}. The magnitude of $v_c$ depends on $\sigma$ in a non-trivial way. For $v_0< 0.86$ the critical velocity $v_c$ can be well approximated by the function $v_c(\sigma) \approx 0.0036339 + 0.589746 \, \sigma^{3.7}$. However, for greater values the dependence is much more complex due to the presence of the resonance phenomenon.

\vspace{0.2cm}

\begin{figure}[h]
\centerline{\includegraphics[height=3.5cm]{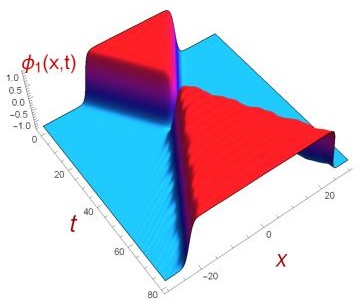} \hspace{1.3cm} \includegraphics[height=3.5cm]{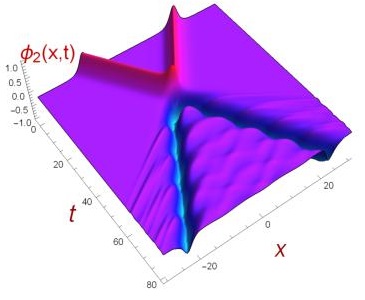}}
\caption{\small Evolution of the first (left) and second (right) component in the $K^{(q,\lambda)}$-$K^{(-q,\lambda)}$ collision with impact velocity $v_0=0.5$ for the model parameter $\sigma=0.72$. The kink and its own antikink collide and mutate into its $\phi_2$-mirror symmetric partners (\textit{one-bounce transmutation regime}).}
\end{figure}

\noindent An extreme type of kink scattering event included in this scenario emerges for large values of the parameter $\sigma$ and collision velocities, see Figure 11. In this case a great amount of the kinetic energy (stored in the zero mode of each traveling kink) is transferred to the transversal discrete eigenmode. The amplitude of this new excited mode is so large that the induced $\phi_2$-fluctuations make the evolving solutions repeatedly jump the potential peak located in the internal plane origin. As a consequence, the scattered kinks periodically mutate into its $\phi_2$-mirror partners, see Figure 11. The radiation emission can decrease the amplitude of this mode and stop this transmutation sequence.

\vspace{0.2cm}

\begin{figure}[h]
\centerline{\includegraphics[height=3.5cm]{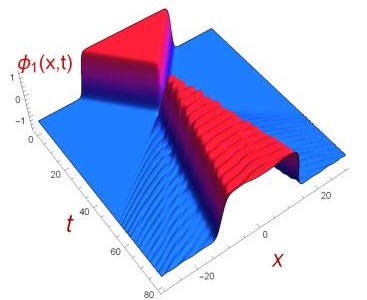} \hspace{1.3cm} \includegraphics[height=3.5cm]{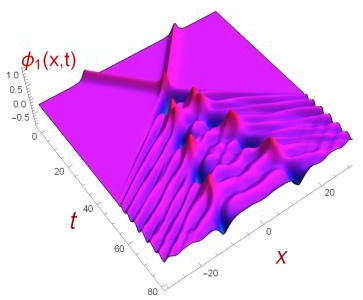}}
\caption{\small Evolution of the first (left) and second component (right) in the $K^{(q,\lambda)}$-$K^{(-q,\lambda)}$ collision with impact velocity $v_0=0.65$ for the model parameter $\sigma=0.94$. Here, the excitation of the $\phi_2$-fluctuations after the kink collision is so large that provokes the successive change between each topological defects and its $\phi_2$-mirror partner.}
\end{figure}

\vspace{0.2cm}

\noindent -- (2) On the other hand, for small enough values of the initial velocity $v_0$ the extended particles described by the solutions $K^{(q,\lambda)}(x)$ and $K^{(-q,\lambda)}(x)$ get trapped in a long lasting bound state (bion). In this process the kink $K^{(q,\lambda)}(x)$ and the antikink $K^{(-q,\lambda)}(x)$ approach each other and collide. The impact turns these solutions into its symmetric partners $K^{(q,-\lambda)}(x)$ and $K^{(-q,-\lambda)}(x)$, which live in the opposite branch of the elliptical orbit (\ref{ellipse}). These new lumps move away a certain distance, but later they attract each other again. The transformed kinks approach, collide and change into the original pair of solutions, the kink $K^{(q,\lambda)}(x)$ and the antikink $K^{(-q,\lambda)}(x)$. The reborn kinks bounce back and move apart until they attract again. This process is repeated over and over emitting a decreasing amount of radiation in every impact. In Figure 12 this scattering event has been plotted for the values $\sigma=0.72$ and $v_0=15$. The region $(\sigma,v_0)$ where this behavior manifests will be referred to as the \textit{bion formation regime}.

\vspace{0.2cm}

\begin{figure}[h]
\centerline{\includegraphics[height=3.5cm]{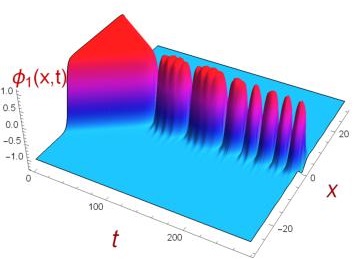} \hspace{1.3cm} \includegraphics[height=3.5cm]{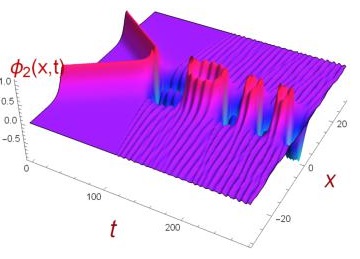}}
\caption{\small Evolution of the first (left) and second component (right) in the $K^{(q,\lambda)}$-$K^{(-q,\lambda)}$ collision with impact velocity $v_0=0.15$ for the model parameter $\sigma=0.72$. The kink and its own antikink collide and form a bound state where a transmutation occurs after every kink collision (\textit{bion formation regime}).}
\end{figure}

\noindent -- (3) For $\sigma \geq 0.68$ the resonant energy transfer mechanism arises in this type of scattering events. An energy exchange between the kink translational mode and the internal vibrational mode is now possible in each collision and so the chance that the kink and antikink escape after colliding a finite number of times. The presence of resonance windows starts timidly for parameter values close to $0.68$ but the effect is accentuated for greater values of $\sigma$. For example, a complex pattern of resonance windows can be observed for $\sigma=0.94$, see Figure 13. The number of bounces suffered by the kinks before escaping is pointed out in this Figure.
\begin{figure}[h]
\centerline{\includegraphics[width=10cm]{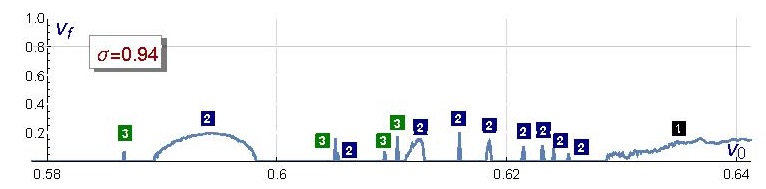} }
\caption{\small Final kink velocity as a function of the initial velocity $v_0\in (0.58,0.64)$ for the coupling constant $\sigma=0.94$ showing the resonance window structure.}
\end{figure}

\noindent It is assumed that other $n$-bounce windows with narrower than our search step $\Delta v_0=10^{-4}$ widths can exist. Indeed, the final scattered kink configuration depends on the number of collisions. In every collision the kinks are transmuted into its $\phi_2$-reflected kinks. As a result, if $N$ is odd the scattering process is characterized by the process (\ref{pheno1}) whereas if $N$ is even a global kink reflection
\[
K^{(q,\lambda)} (v_0)\cup K^{(-q,\lambda)} (-v_0) \rightarrow K^{*(q,\lambda)} (-v_f)\cup K^{*(-q,\lambda)} (v_1) + {\rm radiation}
\]
is found. These phenomena are illustrated in Figure 14, for the parameter $\sigma=0.72$ and $v_0=0.01603$.

\vspace{0.2cm}

\begin{figure}[h]
\centerline{\includegraphics[height=3.5cm]{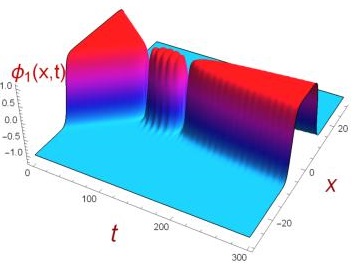} \hspace{1.3cm} \includegraphics[height=3.5cm]{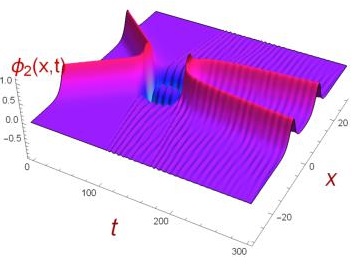}}
\caption{\small Evolution of the first (left) and second component (right) in the two-bounce $K^{(q,\lambda)}$-$K^{(-q,\lambda)}$ collision with impact velocity $v_0=0.1603$ for the model parameter $\sigma=0.72$. The kink and the antikink collide twice before escaping (\textit{resonance regime}).}
\end{figure}

The presence of this resonance scheme can be justified by the existence of the discrete eigenvalue $\nu_1^2$ in the small kink fluctuation operator spectrum. The isolation of this discrete eigenvalue in the spectrum strengthens the resonance mechanism for large values of $\sigma$. On the other hand, for small values of $\sigma$ the continuous spectrum dominates the behaviour of the kink collision and the resonance windows disappear. The MSTB model with $\sigma=1$ restores the $\phi^4$ model resonance window arrangement because in this case the second component of the $K^{(q,\lambda)}(x)$-kinks vanishes and the evolution equations (\ref{klein1}) and (\ref{klein2}) do not change this circumstance.

\subsection{$K_{\rm static}^{(q,\lambda)}(x)- K_{\rm static}^{(-q,-\lambda)}(x)$ scattering processes}

In this Section the study of the collisions between a kink $K^{(q,\lambda)}$ and the antikink $K^{(-q,-\lambda)}$ will be addressed. The $K^{(q,1)}(x)$-solution describes a semi-ellipse orbit confined in the $\phi_2>0$-semi-plane whereas the $K^{(-q,-1)}$-solitary wave survives in the semi-plane $\phi_2<0$, see Figure 4. Again, we are interested in cataloguing the distinct scattering events which are possible in this scenario. This task can be systematized by analyzing the dependence of the final velocity of the scattered kinks as a function of the initial collision velocity and the coupling constant $\sigma$. In this situation, the initial configuration consists of two well separated boosted static kinks
\begin{equation}
K^{(q,\lambda)}(x-x_0,t;v_0) \cup K^{(-q,-\lambda)}(x+x_0,t;-v_0) \label{conca}
\end{equation}
which are pushed together with collision velocity $v_0$. The concatenation (\ref{conca}) describes a loop starting and ending at $A_\mp$ that surrounds the local maxima of the MSTB potential located at the origin $(\phi_1,\phi_2)=(0,0)$, see Figure 4. This loop configuration governs the behaviour of the $K^{(q,\lambda)}-K^{(-q,-\lambda)}$ scattering processes.

\begin{figure}[h]
\centerline{\begin{tabular}{c} \includegraphics[width=5cm]{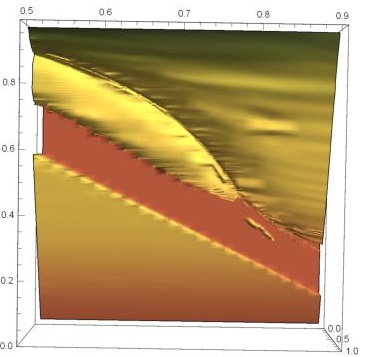} \end{tabular}\hspace{1cm}\begin{tabular}{c} \includegraphics[width=6cm]{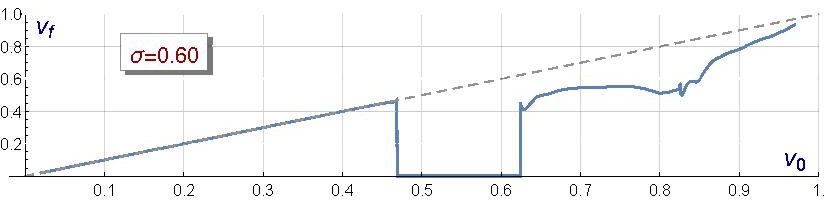} \\
\includegraphics[width=6cm]{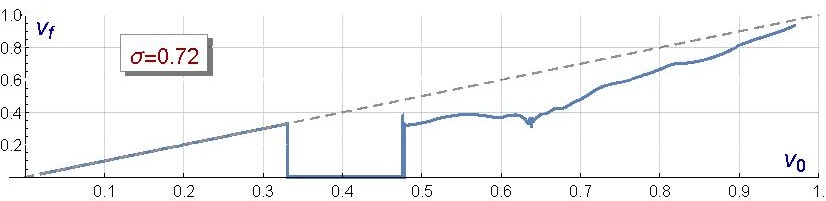} \\
\includegraphics[width=6cm]{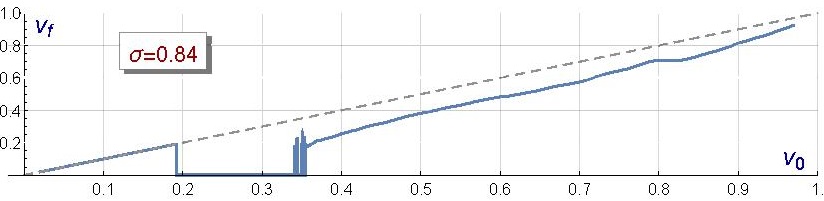} \end{tabular}}
\caption{\small Final kink velocity as a function of the initial velocity $v_0\in (0,1)$ and the model parameter $\sigma\in (0.5,1)$ for the $K^{(q,\lambda)}$-$K^{(-q,-\lambda)}$ collisions (left, top view). Plane sections of the 3D graphics for the values $\sigma=0.60,0.72$ and $0.84$ (right). Zero final velocity indicates mutual kink annihilation.}
\end{figure}

The data found in the numerical analysis are displayed in Figure 15 (left), where the dependence of the final velocity $v_f$ of the scattered kinks on the initial velocity $v_0$ and the parameter $\sigma$ is plotted. Three plane sections of this 3D graphics are included in Figure 15 (right).
The particular behavior of $v_f$ with respect to $v_0$ is shown for particular values of $\sigma=0.60, 0.72$ and $0.84$. The orography presented in Figure 15 (left) allows to distinguish five types of initial velocity regimes, which are described in the following points:

\vspace{0.2cm}

\noindent -- (1)  For low enough collision velocities $v_0$ the kink scattering is almost elastic for any value of the coupling constant $\sigma$. This process, symbolically represented as
\[
K^{(q,\lambda)} (v_0)\cup K^{(-q,-\lambda)} (-v_0) \rightarrow K^{(q,\lambda)} (-v_0)\cup K^{(-q,-\lambda)} (v_0) \hspace{0.5cm} ,
\]
is illustrated in Figure 16, where the evolution of the kink components is displayed. Here, the kinks approach each other with initial velocity $v_0$, collide, bounce back and move away approximately with the same speed. This set of initial velocities and coupling constants, where the two involved kinks are reflected, will be named as the \textit{elastic reflection regime}. The upper boundary of this regime implies a practically linear dependence on the parameter $\sigma$. Indeed, if we denote $v_1(\sigma)=1.16073 - 1.15325 \, \sigma$ then the region $0<v_0 < v_1(\sigma)$ approximately delimitates this domain, as we can see in the Figure 15. In this regime the kink-antikink impact slightly perturbs the simple closed orbit of the initial configuration (\ref{conca}) by introducing fluctuations along the $\phi_1$ and $\phi_2$ components. However, the evolution of this kink-composite loop preserves the simplicity and closeness of the original configuration (\ref{conca}) because the evolving multikink solution is unable to jump the potential maximum.

\vspace{0.2cm}

\begin{figure}[h]
\centerline{\includegraphics[height=3.5cm]{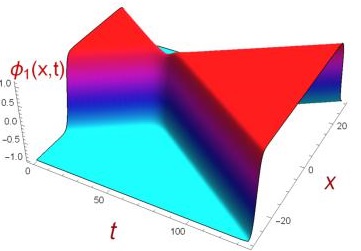} \hspace{1.3cm} \includegraphics[height=3.5cm]{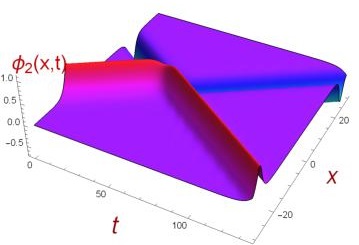}}
\caption{\small Evolution of the first (left) and second component (right) in the $K^{(q,\lambda)}$-$K^{(-q,-\lambda)}$ collision with impact velocity $v_0=0.3$ for the model parameter $\sigma=0.72$. The kink $K^{(q,\lambda)} (v_0)$ and the antikink $K^{(-q,-\lambda)} (-v_0)$ (of different type) collide and elastically reflect (\textit{elastic reflection regime}).}
\end{figure}

\noindent -- (2) For initial velocities $v_0$ ranged in a band of values greater than those defining the previous regime, the kinks mutually annihilate almost instantaneously after the creation of a short-living bound state (bion), see Figure 17. Here, the kink impact is followed by a strong radiation emission. The final configuration consists of traveling plane waves around the vacuum $A_-=-1$. This can be observed in Figure 17 (left) by the lack of red hues in the first component of the evolving solution. This type of processes is characterized as
\[
K^{(q,\lambda)} (v_0)\cup K^{(-q,-\lambda)} (-v_0) \rightarrow {\rm radiation} \hspace{0.3cm} .
\]
For these events the $\phi_1$-perturbations provoked by the collision are strong enough to make the composite kink orbit (\ref{conca}) jump the potential peak located at $(\phi_1,\phi_2)=(0,0)$. This process involves a kinetic energy loss in form of radiation emission and internal mode excitations. This energy loss prevents the evolving solution from returning to the original loop configuration. Consequently, kink annihilation takes place and a radiation vestige remains, see Figure 17. The set of collision velocity windows where these events happen will be referred to as the \textit{annihilation regime}. It is approximately confined in the band $v_1(\sigma) < v_0 < v_2(\sigma)$ where $v_2(\sigma)= 1.29566 - 1.1214 \, \sigma$. However, for large enough values of $\sigma$ some small regions must be excluded in this regime because kink and antikink manage to escape due to a resonance mechanism, see Figure 15. This situation will be discussed later.

\vspace{0.2cm}

\begin{figure}[h]
\centerline{\includegraphics[height=3.5cm]{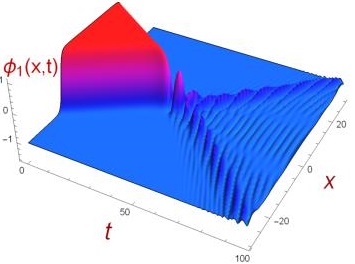} \hspace{1.3cm} \includegraphics[height=3.5cm]{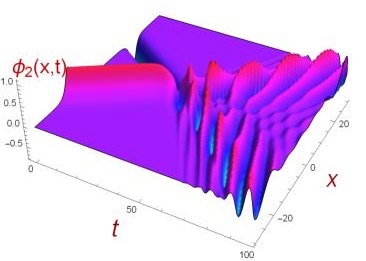}}
\caption{\small Evolution of the first (left) and second component (right) in the $K^{(q,\lambda)}$-$K^{(-q,-\lambda)}$ collision with impact velocity $v_0=0.4$ for the model parameter $\sigma=0.72$. The kink $K^{(q,\lambda)} (v_0)$ and the antikink $K^{(-q,-\lambda)} (-v_0)$ (of different type) collide and mutually annihilate (\textit{annihilation regime}).}
\end{figure}

\noindent -- (3) Another remarkable feature in Figure 15 (left) is the presence of a curve in the $(v_0,\sigma)$-space where the final velocity of the scattered topological defects substantially drops with respect to the its neighbourhood. Therefore, this curve determines the loci of quasiresonances. It is approximately interpolated by the expression $v_q(\sigma)\approx 0.945028 - 1.757 \, \sigma^{5.3}$ for $0.5\leq \sigma \leq 0.8$. An heuristic explanation of the presence of this quasiresonance curve based on the orbit dynamics is as follows: the quasiresonances arise when the kinks $K^{(q,\lambda)}$ and $K^{(-q,-\lambda)}$ (pushed together with velocity $v_q$) evolves to a configuration close to the metastable ${\cal K}^{(q)}$-${\cal K}^{(-q)}$ configuration (with orbit $\phi_2=0$) after the kink-antikink impact. This scenario allows to approximately compute the expression of this curve by using an energetic argument. The original configuration carries a total energy $2E[K_{\rm static}^{(q,\lambda)}(x)]/\sqrt{1-v_0^2}$. A part of the kinetic energy, which can be assessed as $2\rho \{ E[K_{\rm static}^{(q,\lambda)}(x)]/\sqrt{1-v_0^2}-E[K_{\rm static}^{(q,\lambda)}(x)] \}$ with $0<\rho<1$, is lost by radiation emission or vibrational mode excitation. The ${\cal K}^{(q)}$-${\cal K}^{(-q)}$ configuration energy is given by $2 E[{\cal K}_{\rm static}^{(q)}(x)]$. An energy balance leads to the expression
\[
v_q(\sigma)\approx \frac{(1-\sigma) \sqrt{2+\sigma} \, \sqrt{2+\sigma(1-2\rho)(3-\sigma^2)}}{2-3\rho \sigma+\rho \sigma^3}
\]
This expression is a good approximation to the quasiresonance curve for $\rho=0.6$ and this supports the previous interpretation. Besides, this curve is the upper boundary of the \textit{transmutation regime}, defined by the condition $v_2(q)\leq v_0 \leq v_q(\sigma)$. For this regime the $K^{(q,\lambda)}$ and $K^{(-q,-\lambda)}$ kinks collide and emerge as its corresponding antikinks after the impact, see Figure 18. This type of events is symbolically represented as
\begin{equation}
K^{(q,\lambda)} (v_0)\cup K^{(-q,-\lambda)} (-v_0) \rightarrow K{}^{*(q,-\lambda)} (-v_1)\cup K{}^{*(-q,\lambda)}  (v_1) +{\rm radiation} \label{mutation}
\end{equation}
with $v_1<v_0$. The excitation of internal modes has been symbolized by the asterisk superscript in (\ref{mutation}). Another possible interpretation of the previous event is that the kink $K^{(q,\lambda)}$ and the antikink $K^{(-q,-\lambda)}$ collide and reflect exchanging the charge $\lambda$. The collision velocity $v_0$ in this regime is large enough to make the $K^{(q,\lambda)}$-$K^{(-q,-\lambda)}$ solution overcome the potential barrier twice, returning to the original loop configuration. However, the $\phi_2$-fluctuations produced by the kink collision flip the elliptic orbit branches of the original loop configuration (\ref{conca}) with respect to the $\phi_1$ axis. This flip transforms the original kinks into the $K{}^{*(q,-\lambda)} (-v_1)\cup K{}^{*(-q,\lambda)} (v_1)$ configuration.

\begin{figure}[h]
\centerline{\includegraphics[height=3.5cm]{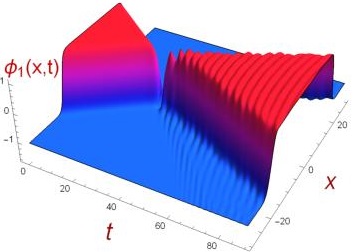} \hspace{1.3cm} \includegraphics[height=3.5cm]{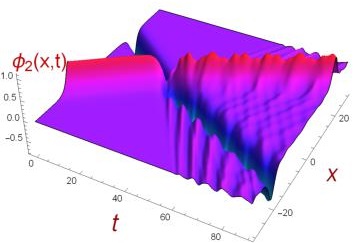}}
\caption{\small Evolution of the first (left) and second component (right) in the $K^{(q,\lambda)}$-$K^{(-q,-\lambda)}$ collision with impact velocity $v_0=0.5$ for the model parameter $\sigma=0.72$. The kink $K^{(q,\lambda)} (v_0)$ and the antikink $K^{(-q,-\lambda)} (-v_0)$ (of different type) collide and mutate into its $\phi_2$-mirror partners (\textit{transmutation regime}).}
\end{figure}

\noindent -- (4) The following distinguished domain corresponds to initial velocities greater than the values of the rest of regimes, that is, for $v_0 > v_q(\sigma)$ and $v_0>v_2(\sigma)$. The scattering processes in this case can be described by the relation
\[
K^{(q,\lambda)} (v_0)\cup K^{(-q,-\lambda)} (-v_0) \rightarrow K^{*(q,\lambda)} (-v_1)\cup K^{*(-q,-\lambda)} (v_1) + {\rm radiation}
\]
with $v_1<v_0$. The final result can be summarized as a non-elastic kink reflection. However, in contrast to the \textit{elastic reflection regime} the kinks in this case suffer several transformations before reaching its ultimate configuration. The loop dynamics follows a similar pattern to the previous regime but now the $\phi_2$-fluctuations induced by the kink collision provoke a double flip in the elliptic orbit branches. This implies that the existence of the $K{}^{*(q,-\lambda)} \cup K{}^{*(-q,\lambda)}$ configuration is ephemeral and the original kink configuration is restored although with excited internal vibrational modes and radiation emission.
The term \textit{inelastic reflection regime} will be coined to name this domain.


\vspace{0.2cm}

\noindent -- (5) As previously mentioned for some ranges of $v_0$ and $\sigma$ a resonant energy transfer mechanism is triggered, which implies that the kinks collide and bounce back a finite number $N$ of times before recovering the kinetic energy necessary to escape. The final result, however, depends on the number of collisions. In every collision the kinks are transmuted into its antikinks. As a result, if $N$ is odd the scattering process is characterized as
\[
K^{(q,\lambda)} (v_0)\cup K^{(-q,-\lambda)} (-v_0) \rightarrow K{}^{*(q,-\lambda)} (-v_1)\cup K{}^{*(-q,\lambda)}  (v_1) +{\rm radiation}
\]
whereas if $N$ is even the scattering events
\[
K^{(q,\lambda)} (v_0)\cup K^{(-q,-\lambda)} (-v_0) \rightarrow K^{*(q,\lambda)} (-v_1)\cup K^{*(-q,-\lambda)} (v_1) + {\rm radiation}
\]
are found. These phenomena are illustrated in Figure 19.

\vspace{0.2cm}

\begin{figure}[h]
\centerline{\includegraphics[height=3.5cm]{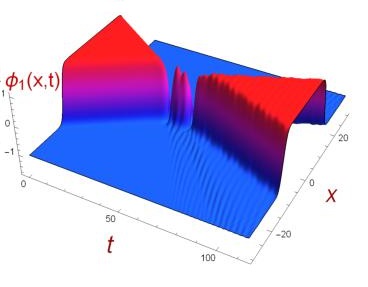} \hspace{1.3cm} \includegraphics[height=3.5cm]{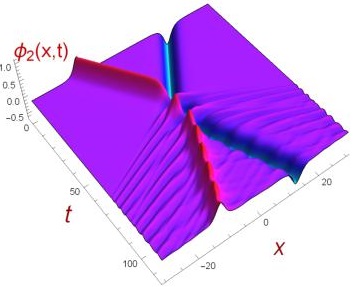}}
\caption{\small Evolution of the first (left) and second component (right) in the $K^{(q,\lambda)}$-$K^{(-q,-\lambda)}$ collision with impact velocity $v_0=0.344$ for the model parameter $\sigma=0.84$. The kink $K^{(q,\lambda)} (v_0)$ and the antikink $K^{(-q,-\lambda)} (-v_0)$ (of different type) collide twice before escaping (\textit{resonance regime}).}
\end{figure}

\section{Conclusions and further comments}

Kink dynamics in the one-parametric family of relativistic $(1+1)$-dimensional two scalar field theories known as MSTB models has been thoroughly explored. Increasing the number of fields in the theory, henceforth enriching the number and types of kinks in a model, enlarges the variety of possible scattering processes. In particular, the presence of two different stable topological kinks doubles the number of kink-antikink scattering events. The kink can be obliged to collide with its own antikink or with the antikink of the other stable topological kink. The output of these processes depends on the collision velocity $v_0$ and the coupling constant $\sigma$ of the model. The domains in the $(v_0,\sigma)$-plane where different types of scattering events take place have been established.

In the $K^{(q,\lambda)}-K^{(-q,\lambda)}$ scattering a one-bounce transmutation regime (where the kinks are converted into its $\phi_2$-mirror symmetric partners $K^{(q,-\lambda)}-K^{(-q,-\lambda)}$ after the collision) and a bion formation regime (where the kink and the antikink collide and bounce back over and over mutating into its symmetric solutions in every impact) are found. The previous pattern is general except for tiny regions where the resonant energy transfer mechanism is turned on. In this situation the bion state is broken after a finite number $N$ of collisions. The final scattered kinks depends on the parity of $N$.

The second class of scattering events in this framework corresponds to the $K^{(q,\lambda)}-K^{(-q,-\lambda)}$ collisions. Now, the range of events is wider. An elastic reflection regime (where kink and antikink elastically reflect), an annihilation regime (where kink and antikink mutually annihilate), a transmutation regime (where kink and antikink mutate into its $\phi_2$-mirror partners) and a ineslastic reflection regime (where the solutions reflect with energy loss) are found. Resonant windows are also found in this context.

The large variety of kink scattering events found in the one-parametric family of MSTB models shows that the study of the kink collisions in $N\geq 2$ scalar field theory models can provide new insights in the broad topic of one-dimensional topological defect (domain walls in $3D$) dynamics. It would be worthwhile, thus, to analyse kink collisions in others models involving several scalar fields.Continuous models of spin chains describing
ferromagnetic or antiferromagnetic phases are particularly interesting in this direction. In Reference \cite{Haldane1983} Haldane showed that kinks
breaking the $O(3)$ symmetry in the non-linear sigma-model with target space a $2D$-sphere are important in characterizing some topological phase
in anti-ferromagnetic materials. The full variety of kinks in this effective model was analyzed by me and my collaborators in the work \cite{Alonso2008b}). The surprising result is that the kink variety in the massivenon-linear sigma model is identical to the kink variety in the MSTB model. There exists a discrete set of stable and unstable topological kinks and one-parametric families of unstable non-topological kinks.
Moreover, in \cite{Alonso2009} the semi-classical corrections to the masses of topological kinks where computed using heat kernel/zeta function regularization methods. It is therefore tantalizing to study kink scattering processes in the massive non-linear sigma  model in the hope of grasping deeper understanding of topological phases in antiferromagnetic materials.

\section*{Acknowledgments}

The author acknowledges the Spanish Ministerio de Econom\'{\i}a y Competitividad for financial support under grant MTM2014-57129-C2-1-P. They are also grateful to the Junta de Castilla y Le\'on for financial help under grant VA057U16.

\end{document}